\global\def\draftcontrol{0}

   \def\versionno{ superpotential }

\catcode`\@=11

\newcommand\makepapertitle{\par

  \begingroup
    \renewcommand\thefootnote{\@fnsymbol\c@footnote}%
 \newpage
     \global\@topnum\z@   
     \@makepapertitle
     \thispagestyle{empty}\@thanks
  \endgroup
  \setcounter{footnote}{0}%
  \global\let\thanks\relax
  \global\let\makepapertitle\relax
  \global\let\@makepapertitle\relax
  \global\let\@thanks\@empty
  \global\let\@author\@empty
  \global\let\@date\@empty
  \global\let\@title\@empty
  \global\let\title\relax
  \global\let\author\relax
  \global\let\date\relax
  \global\let\and\relax
  \def\version{\let\version\@version\@gobble}
}
\def\@makepapertitle{%
  \newpage
   \ifnum\draftcontrol=1 {}
   \version\versionno
   \vskip 5.5em%
   \else
   \hfill\hbox to 3.5cm {\parbox{5cm}{\@pubnum}\hss}%
   \vskip 6.5em%
   \fi
   \begin{center}%
   \let \footnote \thanks
      {\hskip -0\textwidth \hbox to 1\textwidth%
        {\centerline{\Large\bf{\noindent%
    \parbox[t]{1.3\textwidth}{\begin{center}\@title\end{center}}}}}}%
     \vskip 1.5em%
     {\normalsize
       \lineskip .5em%
       \begin{tabular}[t]{c}%
         \@author
       \end{tabular}\par}%
     \vskip 1.5em%
     {\@bstract}%
     \end{center}%
     \vfill
     \@date%
     \vskip 1.5em%
   \par
}

\gdef\@pubnum{}
\def\pubnum#1{%
  \gdef\@pubnum{#1}}

\gdef\@bstract{}
\def\Abstract#1{%
  \gdef\@bstract{%
   \parbox{\textwidth-0pc}{%
   \centerline{\bf Abstract}\penalty1000
   \noindent
   \renewcommand\baselinestretch{1.0}
   {#1}}}
}

\gdef\@email{}
\def\email#1{%
   \gdef\@email{%
   Email: {\tt #1}}
}

\def\ps@paper{\let\@mkboth\@gobbletwo%
     \ifnum\draftcontrol=1
        \def\@oddfoot{\hbox to \textwidth{\tiny \versionno \hfil\tiny\draftdate}%
        \hskip -\textwidth \hbox to \textwidth{\hfil\rm\thepage\hfil}}%
     \else\def\@oddfoot{\hbox to \textwidth{\hfil\rm\thepage\hfil}}
     \fi
     \let\@evenfoot\@oddfoot
}

\def\body{\clearpage
          \pagestyle{paper}
        }

\def\@version#1{\ifnum\draftcontrol=1
\typeout{}\typeout{#1}\typeout{}
\vskip3mm\centerline{\hbox{\fbox{\normalsize{\tt DRAFT -- #1 -- }
                   {\draftdate}}}}\vskip3mm
\fi}
\let\version\@version
\long\def\eqlabel#1{\ifnum\draftcontrol=1
                    \tag@false  
                    \tag*{(\theequation) \hbox to -0.2cm{\hspace{0cm}\small{#1}\hss}}
                    \refstepcounter{equation}
                    \edef\@currentlabel{\theequation}
                    \ltx@label{#1}
                    \else
                    \label{#1}
                    \fi
                    }
\let\st@bibitem\@bibitem
\let\st@lbibitem\@lbibitem
\ifnum\draftcontrol=1
  \def\@bibitem#1{%
    \st@bibitem{#1}\a@@label{#1}\ignorespaces}
  \def\@lbibitem[#1]#2{%
    \st@lbibitem[#1]{#2}\a@@label{#2}\ignorespaces}
  \def\a@@label#1{%
    \gdef\a@lab{\smash{\normalfont\small#1}}
    \ifvmode
      \if@inlabel
        \global\setbox\@labels\hbox{%
          \llap{\a@lab\let\a@lab\relax
                \kern\@totalleftmargin\kern\marginparsep}%
          \box\@labels}%
      \fi
    \fi}
\fi

\documentclass[12pt,letterpaper]{article}
\usepackage{amsmath}
\usepackage{amsmath}
\usepackage{amsmath}
\usepackage{amssymb}
\usepackage{amssymb}
\usepackage{amssymb}
\usepackage{amsmath,bm,amsfonts,amssymb,array,calc,amsthm,rotating,cite}
\usepackage{epsfig,psfrag}
\usepackage{graphicx}
\usepackage{color}
\usepackage[colorlinks=true]{hyperref}
\usepackage[all]{xy}

\renewcommand\baselinestretch{1.25}
\renewcommand\section{\@startsection {section}{1}{\z@}%
                                   {-3.5ex \@plus -1ex \@minus -.2ex}%
                                   {2.3ex \@plus.2ex}%
                                   {\normalfont\large\bfseries}}
\renewcommand\subsection{\@startsection{subsection}{2}{\z@}%
                                   {-3.25ex\@plus -1ex \@minus -.2ex}%
                                   {1.5ex \@plus .2ex}%
                                   {\normalfont\normalsize\bfseries}}
\renewcommand\subsubsection{\@startsection{subsubsection}{3}{\z@}%
                                   {-3.25ex\@plus -1ex \@minus -.2ex}%
                                   {1.5ex \@plus .2ex}%
                                   {\normalfont\normalsize\it}}
\renewcommand\paragraph{\@startsection{paragraph}{4}{\z@}%
                                   {-3.25ex\@plus -1ex \@minus -.2ex}%
                                   {1.5ex \@plus .2ex}%
                                   {\normalfont\normalsize\bf}}
\renewcommand\subparagraph{\@startsection{subparagraph}{5}{\z@}%
                                   {-1.25ex\@plus -1ex \@minus -.2ex}%
                                   {0ex \@plus .2ex}%
                                   {\normalfont\normalsize\it}}

\numberwithin{equation}{section}

\renewcommand*\l@section[2]{%
  \ifnum \c@tocdepth >\z@
    \addpenalty\@secpenalty
    \addvspace{.5em \@plus\p@}%
    \setlength\@tempdima{1.5em}%
    \begingroup
      \parindent \z@ \rightskip \@pnumwidth
      \parfillskip -\@pnumwidth
      \leavevmode \bfseries
      \advance\leftskip\@tempdima
      \hskip -\leftskip
      #1\nobreak\hfil \nobreak\hb@xt@\@pnumwidth{\hss #2}\par
    \endgroup
  \fi}
\renewcommand*\l@subsection{\addvspace{.0em \@plus\p@}\@dottedtocline{2}{1.5em}{2.3em}}
\renewcommand*\l@subsubsection{\addvspace{-.2em \@plus\p@}\@dottedtocline{3}{3.8em}{3.2em}}

\definecolor{refcol}{rgb}{0.0,0.0,0.2}
\definecolor{eqcol}{rgb}{.2,0,0}
\definecolor{purple}{cmyk}{0,1,0,0}

\gdef\@citecolor{refcol} \gdef\@linkcolor{eqcol}
\gdef\@urlcolor{refcol}
\def\colorlinkspurple{\gdef\@urlcolor{purple}}
\def\colorlinksblue{\gdef\@urlcolor{blue}}
\def\colorlinksred{\gdef\@urlcolor{red}}

\def\revise#1       {\raisebox{-0em}{\rule{3pt}{1em}}%
                     \marginpar{\raisebox{.5em}{\vrule width3pt\
                     \vrule width0pt height 0pt depth0.5em
                     \hbox to 0cm{\hspace{0cm}{%
                     \parbox[t]{4em}{\raggedright\footnotesize{#1}}}\hss}}}}

\catcode`\@=12
\newcommand{\bqa}{\begin{eqnarray}}
\newcommand{\eqa}{\end{eqnarray}}
\begin{document}

\title{
D-brane Superpotentials and Ooguri-Vafa Invariants of Compact Calabi-Yau Threefolds}

\author{
Feng-Jun Xu,~~~Fu-Zhong Yang\footnote{Corresponding author~~~
E-mail: fzyang@gucas.ac.cn} \\[0.2cm]
\it College of Physical Sciences, Graduate University of Chinese Academy of Sciences\\
\it   YuQuan Road 19A, Beijing 100049, China}

\Abstract{~~~~~We calculate the D-brane superpotentials for two non-Fermat type compact Calabi-Yau manifolds which are
 the hypersurface of degree 14 in the weighed projective space P(1,1,2,3,7) and 
the hypersurface of degree 8 in the weighed projective space P(1,1,1,2,3) in type II string theory£¬ respectively. 
 By constructing the open-closed mirror maps, we also compute the Ooguri-Vafa invariants, which are related to the open Gromov-Witten invariants.}

\makepapertitle

\body

\version\versionno

\vskip 1em

\newpage

\section{Introduction}

~~~~When considering certain $\mathcal{N}=1$ supersymmetric string
compactifications of type II string theories with space-filling
D-branes, superpotential is an important quantity due to its
BPS-property, which is exactly solvable. Superpotential is both
important in physics and mathematics. The physical interest
 is served by facts that superpotential is also known the holomorphic F-term in effective
lagrangian. The mathematical application is related to the
non-perturbative stringy geometry. Similar to the
  Closed string theory, there is a geometry parameterizing the moduli
space, namely N=1 special geometry. From the viewpoint of
Mathematics, the presence of a superpotential describe an
obstruction to continues deformation of moduli space. And even more
surprise results is in enumerative geometry.  The superpotentials of
A-model at large radius region count the disk invariants
\cite{Aganagic:2000gs} which are related to open Gromov-Witten
invariants \cite{fukaya,L,KL,Jinzenji,Fang}.

   The $N=1,~d=4$ superpotential term can be computed by the open topological string amplitudes
$\mathcal{F}_{g,h}$
 of the A-model as follows
 \bqa
 \eqlabel{top}
h\int
d^4xd^2\theta\mathcal{F}_{g,h}(\mathcal{G}^2)^g(\mathcal{F}^2)^{h-1}
 \eqa
where $\mathcal{G}$ is the gravitational chiral superfield and
$\mathcal{F}$ is the gauge chiral superfield. The formula
\eqref{top} at $g=0,~h=1$ leads to in F-terms of $N=1$ supersymetric
theories:
$$\int d^4xd^2\theta \mathcal{W}(\Phi)$$.

  For non-compact Calabi-Yau manifolds, the refs \cite{Aganagic:2000gs,Kachru:2000an,Kachru:2000ih,Aganagic:2001nx,Lerche:2002yw,Lerche:2002ck,Aganagic:2003db} studied the open-closed mirror
symmetry and its applications. In particular, the work
\cite{Aganagic:2000gs} constructed the classical A-brane geometry
with special Lagrangian submanifold and the work
\cite{Lerche:2002yw,Lerche:2002ck} introduced $N=1$ spacial geometry
and variation of mixed Hodge structure to calculate superpotentials.
Motivated and guided by these works, a progress on compact manifolds
came from
\cite{Walcher:2006rs,Morrison:2007bm,Krefl:2008sj,Knapp:2008uw},
which studied a class of involution branes independent of open
deformation moduli. Furthermore, there appeared some related works
on superpotential for compact Calabi-Yau manifolds depending on
open-closed deformation moduli
\cite{Alim:2009rf,Alim:2009bx,Alim:2010za,Li:2009dz,Jockers:2008pe,Jockers:2009mn,Jockers:2009ti,Alim:2011rp,Walcher:2009uj,Grimm:2008dq,Grimm:2009sy,Grimm:2010gk,Klevers:2011xs,Grimm:2009ef,Baumgartl:2007an,Baumgartl:2008qp,Baumgartl:2010ad,Shimizu:2010us,Fuji:2010uq}.
where the works \cite{Baumgartl:2007an,Baumgartl:2008qp} studied it
by the conformal field theory and matrix factorization, the works
\cite{Jockers:2008pe,Jockers:2009mn} by the direct integration and
the others considered with Hodge theoretic method.

   In this paper, we calculate D-brane superpotentials for compact non-Fermat Calabi-Yau threefolds by open-closed mirror symmetry and generalized GKZ system
   \cite{Batyrev:1993wa,Batyrev:1994hm,Hosono:1993qy,Hosono:1995bm,GKZ}. There exists a
   duality between the type II compactification with brane on the
   threefold and the M/F-theory compactify on the Calabi-Yau
   fourfold without any branes but
   with fluxes
   \cite{Alim:2009bx,Alim:2009rf,Alim:2010za,Mayr:2001xk,Grimm:2009ef,Klevers:2011xs}.
    In the weak decoupling
   limit $g_s\rightarrow0$, the Gukov-Vafa-Witten superpotentials \cite{Gukov:1999ya} $\mathcal{W}_{GVW}$ of
   F-theory compactify on this fourfold agrees with superpotentials $\mathcal{W}$ of Type II
   compactify threefold with branes at lowest order in $g_s$
   \cite{Alim:2009bx,Alim:2010za,Jockers:2009ti,Berglund:2005dm,Grimm:2009ef,Klevers:2011xs}
   \bqa
\mathcal{W}_{GVW}=\mathcal{W}+\mathcal{O}(g_s)+\mathcal{O}(e^{-1/g_s})
   \eqa
Hence, in this limit, we can obtain the flux superpotential
$\mathcal{W}_{GVW}$ from the superpotential $\mathcal{W}$ which will
be given in this paper.

   In Sect. 2 and Sect. 3  we give a
   overview of Superpotentials on Calabi-Yau Threefolds and generalized GKZ system.
   In sect. 4, we concentrates on the applications with two compact non-Fermat Calabi-Yau threefold with two deformation parameters---$X_{14}(1,1,2,3,7)$ and
   $X_{8}(1,1,1,2,3)$. Although the compact Calabi-Yau manifolds are
   available in many works above, the works about non-fermat case are few.
   In these manifolds, We consider superpotential, mirror symmetry and Ooguri-Vafa invariants for D-brane with a single open deformation moduli. Sect. 5 is for
   summary.

\section{Some Known Results}
~~~In this section, we collect some known results on D-brane
superpotenatial in type II string compactification.

\subsection{Superpotentials on Calabi-Yau Threefolds}

 ~~~~Type II compactification theory is
described by an effective $N=1$ supergravity action with non-trivial
superpotentials on the deformation space $\mathcal{M}$ when adding
D-branes and background fluxes. For D5-brane wrapped the whole
Calabi-Yau threefold, the holomorphic Chern-Simons theory
\cite{Witten:1992fb} \bqa
  \mathcal{W}=\int_{X}\Omega^{3,0}\wedge \text{Tr}[A\wedge \bar{\partial}A+\frac{2}{3}A\wedge A\wedge A]
  \eqa
gives the brane superpotential $\mathcal{W}_{brane}$, where $A$ is
the gauge field with gauge group $U(N)$ for $N$ D6-branes. When
reduced dimensionally, the low dimenaional brane superpotentials can
be obtained as \cite{Lerche:2003hs,Aganagic:2000gs}
 \bqa
\mathcal{W}_{brane}
=N_{\nu}\int_{\Gamma^{\nu}}\Omega^{3,0}(z,\hat{z})=\sum_{\nu}N_{\nu}\Pi^{\nu}
 \eqa
 where $\Gamma^{\nu}$ is a special Lagrangian 3-chain and
$(z,\hat{z})$ are closed-string complex structure moduli and D-brane
moduli from open-string sector, respectively.

    The background fluxes $H^{(3)}=H_{RR}^{(3)}+\tau H_{NS}^{(3)}$, which take values in the integer cohomology group $H^3(X,\mathbb{Z})$, also break the supersymmetry $N=2$ to $N=1$.
 The $\tau=C^{(0)}+ie^{-\varphi}$ is the complexified Type IIB coupling field. Its
 contribution to superpotentials is \cite{Mayr:2000hh,Taylor:1999ii}
 \bqa
 \mathcal{W}_{flux}(z)\ =\int_{X}
 H_{RR}^{(3)}\wedge  \Omega^{3,0} = \sum_\alpha N_\alpha\cdot\Pi^\alpha(z)\ ,
 \qquad N_\alpha \in Z.
\eqa

  The contributions of D-brane and background flux (here the NS-flux ignored) give together the general
  form of superpotential as follow\cite{Lerche:2002ck,Lerche:2002yw}
  \bqa
\mathcal{W}(z,\hat{z})=\mathcal{W}_{brane}(z,\hat{z})+\mathcal{W}_{flux}(z)=\sum_{\gamma_{\Sigma}\in
H^3(Z^\ast,\mathcal{H})}N_{\Sigma}\Pi_{\Sigma}(z,\hat{z}) \eqa where
$N_{\Sigma}=n_{\Sigma}+\tau m_{\sigma}$, $\tau$ is the dilaton of
type II string and $\Pi_{\Sigma} $ is a relative periods defined in
a relative cycle $\Gamma\in H_3(X,D)$ whose boundary is wrapped by
D-branes.

   The off-shell
tension of D-branes, $\mathcal{T}(z,\hat{z})$, is equal to the
relative period \cite{Witten:1997ep,Lerche:2002ck,Lerche:2002yw}
\bqa \Pi_{\Sigma}=\int_{\Gamma_{\Sigma}}\Omega(z,\hat{z}) \eqa
 which
measures the difference between the value of on-shell
superpotentials for the two D-brane configurations\bqa
\mathcal{T}(z,\hat{z})=\mathcal{W}(C^+)-\mathcal{W}(C^-) \eqa  with
$\partial \Gamma_{\Sigma}=C^+-C^-$. The  on-shell domain wall
tension is \cite{Alim:2010za} \bqa
T(z)=\mathcal{T}(z,\hat{z})\mid_{\hat{z}=\text{critic points}}\eqa
where the critic points correspond to $\frac{dW}{d\hat{z}}=0$
\cite{Witten:1997ep} and the $C^{\pm}$ is the holomorphic curves at
those critical points. At those critical points, the domain wall
tensions are also known as normal function giving the Abel-Jacobi
invariants \cite{Clemens,
Alim:2010za,Li:2009dz,Morrison:2007bm,Griffiths}

    In A-model interpretation, the superpotential expressed in term of flat coordinates $(t,\hat{t})$ is
the generating function of the Ooguri-Vafa invariants
\cite{Aganagic:2000gs,Ooguri:1999bv,Lerche:2002ck,Alim:2009rf} \bqa
\mathcal{W}(t,\hat{t})=\sum_{\overrightarrow{k},\overrightarrow{m}}G_{\overrightarrow{k},\overrightarrow{m}}q^{d\overrightarrow{k}}\hat{q}^{d\overrightarrow{m}}=\sum_{\overrightarrow{k},\overrightarrow{m}}\sum_{d}n_{\overrightarrow{k},\overrightarrow{m}}\frac{q^{d\overrightarrow{k}}\hat{q}^{d\overrightarrow{m}}}{k^2}
\eqa where $q=e^{2\pi it}$, $\hat{q}=e^{2\pi i\hat{t}}$ and
$n_{\overrightarrow{k},\overrightarrow{m}}$ are Ooguri-Vafa
invariants\cite{Ooguri:1999bv} counting disc instantons in relative
homology class $(\overrightarrow{m},\overrightarrow{k})$, where
$\overrightarrow{m}$ represents the elements of $H_1(D)$ and
$\overrightarrow{k}$ represents an element of $H_2(X)$.
$G_{\overrightarrow{k},\overrightarrow{m}}$ are open Gromov-Witten
invariants. From string world-sheet viewpoint, these terms in the
superpotential represents the contribution from instantons of sphere
and disk.

\subsection{Relative periods and Generalized GKZ system}
~~~~The generalized hypergeometric systems originated from
\cite{GKZ} and have been applied in mirror symmetry
\cite{Hosono:1993qy,Hosono:1995bm,Batyrev:1993wa,Batyrev:1994hm,Hosono}
. The notation is as follows: $(X^{\ast},X)$ is the mirror pair of
compact Calabi-Yau threefold defined as hypersurfaces in toric
ambient spaces $(W^{\ast},W)$, respectively. The generators $l^a$ of
Mori cone of the toric variety
\cite{Fujino,Fujino2,Scaramuzza,Renesse} give rise to the charge
vectors of the gauged linear sigma model (GLSM)\cite{Witten:1993yc}.
$\bigtriangleup$ is a real four dimensional reflexive polyhedron.
$W=P_{\Sigma(\bigtriangleup)}$ is the toric variety with fan
$\Sigma(\bigtriangleup)$ being the set of cones over the faces of
$\bigtriangleup^\ast$. $\bigtriangleup^\ast$ is the dual polyhedron
and $W^\ast$ is the toric variety obtained from
$\Sigma(\bigtriangleup^\ast)$. The enhanced polyhedron
$\underline{\triangle^{\ast}}$ constructed from polyhedron
$\bigtriangleup^\ast$ is associated to $X_4^{\ast}$ on which the
dual F-theory compactify. The three-fold X on B-model side is
defined by $p$ integral points of $\triangle^{\ast}$ as
   the zero locus of the polynomial $P$ in the toric ambient space
   \bqa
P=\sum_{i=0}^{p-1}a_i\prod_{k=0}^{4}X_k^{\nu_{i,k}^{\ast}}
   \eqa
where the $X_k$ are coordinates on an open torus
$(\mathbb{C}^{\ast})^4\in W$ and $a_i$ are complex parameters
related to the complex structure of X. In terms of homogeneous
coordinates $x_j$ on the toric ambient space, it can be rewritten as
\bqa  \eqlabel{constraint}
 P=\sum_{i=0}^{p-1}a_i\prod_{\nu\in
\triangle}x_j^{\langle\nu,\nu_i^{\ast}\rangle+1}. \eqa

  The open-string sector from D-branes can be described by the family
  of hypersurfaces $\mathcal{D}$, which is defined as intersections
$P=0=Q(\mathcal{D})$. In toric variety, the $Q(\mathcal{D})$ can be
defined as \cite{Alim:2009rf,Alim:2010za} \bqa
Q(\mathcal{D})=\sum_{i=p}^{p+p'-1}a_iX_k^{\nu_{i,k}^{\ast}}\eqa
where additional $p'$ vertices $v_i^{\ast}$ correspond with the
monomials in $Q(\mathcal{D})$.

  When considering the dual F-theory
compactification on Four-fold $X_4$, the relevant "Enhanced
polyhedron" consists of extended vertices
\begin{equation}
\eqlabel{four} \underline{\overline{\nu}_i^{\ast}}=\left\{
\begin{aligned}
(\nu_i^{\ast},0)~~i=0,...,p-1 \\
(\nu_i^{\ast},1)~i=p,...,p+p'-1.
\end{aligned}\right.
\end{equation}

   The period integrals can be written as
   \bqa
   \Pi_i=\int_{\gamma_i}\frac{1}{P(a,X)}{\prod_{j=1}^n \frac{d
   X_j}{X_j}}.
   \eqa
   According to the refs.\cite{Batyrev:1993wa,Batyrev:1994hm}, the
   period integrals can be annihilated by differential operators
   \bqa
   \eqlabel{operator}
   \begin{split}
   &\mathcal{L}(l)=\prod_{l_i>0}(\partial_{a_i})^{l_i}-\prod_{l_i<0}(\partial_{a_i})^{l_i}\\
   &\mathcal{Z}_k=\sum_{i=0}^{p-1}\nu_{i,k}^{\ast}\vartheta_i,
   \qquad \mathcal{Z}_0=\sum_{i=0}^{p-1}\vartheta_i-1
   \end{split}
   \eqa
where $\vartheta_i=a_i\partial_{a_i}$. As noted in
refs.\cite{Hosono:1993qy}\cite{Alim:2009bx}, the equations
$\mathcal{Z}_k\Pi(a_i)=0$ reflex the invariance under the torus
action, defining torus invariant algebraic coordinates $z_a$ on the
moduli space of complex structure of $X$ \cite{Alim:2010za}: \bqa
\eqlabel{coordinates} z_a=(-1)^{l_0^{a}}\prod_{i}a_i^{l_i^{a}}\eqa
where $l_a$, ~$a=1,...,h^{2,1}(X)$ is generators of the Mori cone,
one can rewrite the differential operators $\mathcal{L}(l)$ as
\cite{Batyrev:1994hm,Hosono:1993qy,Alim:2010za}
    \bqa
    \mathcal{L}(l)=\prod_{k=1}^{l_0}(\vartheta_0-k)\prod_{l_i>0}\prod_{k=0}^{l_i-1}(\vartheta_i-k)-(-1)^{l_0}z_a\prod_{k=1}^{-l_0}(\vartheta_0-k)\prod_{l_i<0}\prod_{k=0}^{l_i-1}(\vartheta_i-k).
    \eqa

   The solution to the GKZ system can be written as \cite{Batyrev:1994hm,Hosono:1993qy,Alim:2010za} \bqa
B_{l^a}(z^a;\rho)=\sum_{n_1,...,n_N\in Z_0^{+}}\frac{\Gamma(1-\sum_a
l_0^a(n_a+\rho_a))}{\prod_{i>0}\Gamma(1+\sum_a
l_i^a(n_a+\rho_a))}\prod_a z_a^{n_a+\rho_a}.\eqa

 In this paper we consider the family of divisors $\mathcal{D}$ with a single open
 deformation moduli $\hat{z}$ \footnote{In refs \cite{Grimm:2008dq,Grimm:2010gk} they considered another approach which blows up along the curve C and replaces the pair(X,C) with a non-Calabi-Yau manifold $\widehat{X}$.}
  \bqa
    x_1^{b_1}+\hat{z}x_2^{b_2}=0
  \eqa where $b_1,~b_2$ are some appropriate integers. The relative 3-form $\underline{\Omega}:=(\Omega_{X}^{3,0},0)$ and the relative
periods satisfy a set of differential equations
\cite{Lerche:2002ck,Lerche:2002yw,Alim:2009bx,Alim:2010za,Jockers:2008pe}
 \bqa
\mathcal{L}_a(\theta,\hat{\theta})\underline{\Omega}=d\underline{\omega}^{(2,0)}~\Rightarrow~\mathcal{L}_a(\theta,\hat{\theta})\mathcal{T}(z,\hat{z})=0.
\eqa with some corresponding two-form $\underline{\omega}^{(2,0)}$.
The differential operators $\mathcal{L}_a(\theta,\hat{\theta})$ can
be expressed as \cite{Alim:2010za} \bqa
\mathcal{L}_a(\theta,\hat{\theta}):=\mathcal{L}_a^{b}-\mathcal{L}_a^{bd}\hat{\theta}
\eqa for $\mathcal{L}_a^{b}$ acting only on bulk part from closed
sector, $\mathcal{L}_a^{bd}$ on boundary part from open-closed
sector and $\hat{\theta}=\hat{z}\partial_{\hat{z}}$. The explicit
form of these operators will be given in following model. One can
obtain \bqa \eqlabel{sub} 2\pi
i\hat{\theta}\mathcal{T}(z,\hat{z})=\pi(z,\hat{z}) \eqa by reduction
to Nother-Lefshetz locus for only the family of divisors
$\mathcal{D}$ depending on the $\hat{z}$. From above one can obtain
differential equation with the inhomogeneous term $f_a(z)$ at the
critical points \bqa \mathcal{L}_a^{b}T(z)=f_a(z) \eqa and \bqa 2\pi
i f_a(z)=\mathcal{L}_a^{bd}\pi(z,\hat{z})|_{\hat{z}=\text{critic
points}}
 \eqa

\section{Superpotential of Calabi-Yau $X_{14}(1,1,2,3,7)$}
~~~~The $X_{14}(1,1,2,3,7)$ is defined as a degree 14 hypersurface
which is the zero locus of polynomial $P$ \bqa
 P=x_1^{14}+x_2^{14}+x_3^{7}+x_3x_4^{4}+x_5^{2} \eqa

   The GLSM charge vectors for this manifold are
   \cite{Berglund:1995gd}
   \begin{equation}
\begin{tabular}{c|c c c c c c c }
~           & $0$  & $1$ & $2$& $3 $ & $4$ & $5$& $6$  \\\hline
$l_1$ & $0$ & $-2$ & $-2$& $-4$ & $1$ & $0$ & $7$   \\
$l_2$ &$-2$ & $1$& $1$  & $2$& $0$ & $1$ & $-3$
\end{tabular}
\end{equation}

    The mirror manifold is \cite{Berglund:1995gd} $X^\ast$ is $X^\ast=\widehat{X}/H$, where
$\widehat{X}=X_{28}(2,2,3,7,14)$ and
$H=(h_i^k)=\frac{1}{14}(1,13,0,0,0),\frac{1}{2}(1,0,0,0,1)$ which
act by $x_i\rightarrow  \text{exp}(2\pi ih_i^k)x_i$.
   We consider the following curves
  \bqa
 C_{\alpha,\pm}=\{x_4=\xi
  x_5,x_3=0,x_1^{7}x_4+x_2^{7}x_5=0\}, ~\xi^{3}=-1
  \eqa
which are on the family of divisor  \bqa \eqlabel{Divisor1}
Q(\mathcal{D})=x_4^{3}+z_3x_5^{3}\eqa at the critical points
$z_3=1$.

By the generalized GKZ system, the period on the surface
$P=0=Q(\mathcal{D})$ has the form \bqa
\begin{split}
&\pi=\frac{c}{2}B_{(\hat{l}_1,\hat{l}_2)}(u_1,u_2;0,\frac{1}{2})=\sum_{n_1,n_2}\frac{cz_1^{n_1}z_2^{\frac{1}{2}+n_2}\Gamma(3(n_1+\frac{1}{2})+(n_2)+1)}{\Gamma(2+2n_1)\Gamma(1+n_2)^3\Gamma(n_1-2n_2)}\\
&=-\frac{4c}{\pi^{\frac{3}{2}}}\sqrt{u_1}u_2+\mathcal{O}((u_1u_2)^{3/2})
\end{split} \eqa which vanishes at the critical locus $u_2=0$ in
terms of new parameters as \bqa u_1=\frac{-z_1}{z_3}(1-z_3)^2 \qquad
u_2=z_2\eqa and $Z_{1,2}$ are relevant coordinates in the large
complex structure limit defined as \eqref{coordinates} . Following
the \cite{Feng}, the off-shell superpotentials can be obtained by
integrating the $\pi$:
  \bqa
  \mathcal{T}_a^{\pm}(z_1,z_2)=\frac{1}{2\pi
  i}\int\pi(z_3)\frac{dz_3}{z_3},
\eqa with the appropriate integral constants\cite{Alim:2010za},
 the superpotentials can be chosen as  $\mathcal{W}^+=-\mathcal{W}^-$. In this convention, the on-shell
 superpotentials can be obtained as
\bqa 2\mathcal{W}^{+}=\frac{1}{2\pi
 i}\int_{-z_3}^{z_3}\pi(z_3)\frac{dz_3}{z_3},~~~W^{\pm}(z_1,z_2)=\mathcal{W}^{\pm}(z_1,z_2,z_3)|_{z_3=1}
\eqa
 Eventually, The superpotential are
\bqa
\begin{split}
&\mathcal{W}^{\pm}(z_1,z_2,z_3)=\mp\sum_{n_1,n_2}\frac{cz_1^{n_1}z_2^{\frac{1}{2}+n_2}z_3^{\frac{-1-2n_1}{2}}\Gamma(3n_1+n_2+1)}{4\pi(-1+4n_1^2)\Gamma(2+2n_1)\Gamma(1+n_2)^3\Gamma(n_1-2n_2)}\\
&\{(1-2n_1)
2F_1(-\frac{1}{2}-n_1,-2n_1,\frac{1}{2}-n_1;z_3)+(1+2n_1)z_32F_1((\frac{1}{2}-n_1,-2n_1,\frac{3}{2}-n_1;z_3))\}
\end{split}
\eqa and those can divide two parts \bqa
\mathcal{W}^{\pm}(z_1,z_2,z_3)=W^{\pm}(z_1,z_2)+f(z_1,z_2,z_3)\eqa
where the $f(z_1,z_2,z_3)$ are related to the open-string parameter,
$W$ are the on-shell superpotential defined as
 \bqa
W^{\pm}=\mp\frac{c}{8}\sum_{n_1,n_2}\frac{z_1^{n_1}z_2^{\frac{1}{2}+n_2}\Gamma(3(n_1)+(n_2+\frac{1}{2})+1)}{\Gamma(n_2+\frac{1}{2}+1)^3\Gamma(n_1+1)^2\Gamma(n_1-2n_2)}.
 \eqa

The additional GLSM charge vectors corresponding to the divisor
\eqref{Divisor1}
   are
       \begin{equation}
\begin{tabular}{c|c c c c c c c c c c}
 & $0$ & $1$& $2$& $3 $ & $4$ & $5$ & $6$ & $7$&  $8$&\\\hline
$l_4$ & $-1$ & $1$ & $0$& $0$ & $0$ & $0$ &$0$ & $1$ & $-1$
\end{tabular}
\end{equation}
  The classic A-brane in the mirror Calabi-Yau manifold $X^{\ast}$ of X
determined by the additional charge vectors $(0,-1,1,0,0,0,0)$ is a
special Lagrangian submanifold of $X^{\ast}$ defined as
\cite{Aganagic:2000gs,Aganagic:2001nx,Kachru:2000an,Kachru:2000ih,Alim:2009rf}
\bqa -|x_4|+|x_5|=\eta\eqa where $x_i$ are coordinates on
$X^{\ast}$, $\eta$ is a K$\ddot{a}$hler moduli parameter with
$\hat{z}=\epsilon e^{-\eta}$ for a phase $\epsilon$.

\begin{table}[!h]
\def\temptablewidth{1.0\textwidth}
\begin{center}
\begin{tabular*}{\temptablewidth}{@{\extracolsep{\fill}}c|ccccc}
 $d_1\backslash d_2$& 0     &1     &2    &3
&9\\\hline
1     & 2         & 0     & 0   &0   &0    \\
3     & -2     & 0   & 0  &0&0      \\
5     & 10   &-220 &0 &0  &0       \\
7     & -84    &400 &0&0   &0    \\
9      & 858&-1844 &-13500  &0&0  \\
11    &-9878 &-61760&1501528&0&0
     \end{tabular*}
       {\rule{\temptablewidth}{1pt}}
\end{center}
       \end{table}

The flat coordinates in
 A-model at large radius regime are related to the flat coordinates of B-model at large complex structure regime by mirror map $t_i=\frac{\omega_i}{\omega_0},~\omega_i:= D_i^{(1)}
 \omega_0(z,\rho)|_{\rho=0}$
 \bqa
 \begin{split}
 &q_1=z_1-6z_1^2+63z_3
 -866z_1^4+68z_1^3z_2+\mathcal{O}(z^5)\\
 &q_2=z_2+14z_1z_2-7z_1^2z_2+294z_1^3z_2-96z_1^2z_2^2+\mathcal{O}(z^5)
\end{split}
 \eqa

and we can obtain the inverse mirror map in terms of $q_i=e^{2\pi i
t_i}$
 \bqa
\begin{split}
&z_1=q_1+6q_1^2+9q_1^3+56q_1^4-68q_1^3q_2+\mathcal{O}(q^5)\\
 &z_2=q_2-14q_1q_2-119q_1^2q_2-924q_1^3q_2+96q_1^2q_2^2\mathcal{O}(q^5).
\end{split}
 \eqa

 Using the modified multi-cover
formula\cite{Walcher:2006rs,Walcher:2009uj} for this case
  \bqa
\frac{W^{\pm}(z(q))}{w_0(z(q))}=\frac{1}{(2 i\pi)^2}\sum_{k~
 odd}~\sum_{d_2 , d_1 odd\geq
0}n_{d_1,d_2}^{\pm}\frac{q_1^{kd_1}q_2^{kd_2/2}}{k^2}\eqa the
superpotentials $W^\pm$, at the critical points $z_3=1$, give
Ooguri-Vafa invariants $n_{d_1,d_2}$ for the normalization constants
  $c=1$, which are listed in Table.1.

  Another interesting thing which should be mentioned is the
  superpotential $W^{+}$ encode the information of superpotential of
the non-compact geometry $\mathcal{O}(-3)_{\mathbb{P}^2}$ in the
limit of $q_1\rightarrow 0$. This can be shown by $n_{d_1,0}=n_k$,
where $n_k$ is the disc invariants of
$\mathcal{O}(-3)_{\mathbb{P}^2}$ which were studied in work
\cite{Lerche:1996ni}. See more details in appendix of
\cite{Alim:2010za}.

\section{Superpotential of Calabi-Yau $X_8(1,1,1,2,3)$}
~~~~The $X_{8}(1,1,1,2,3)$ is defined as a degree 8 hypersurface
which is the zero locus of polynomial $P$ \bqa P
=x_1^{8}+x_2^{8}+x_3^{8}+x_4^{4}+x_4x_5^{2} \eqa
    The GLSM charge vectors $l_a$ are the generators of the Mori cone as follows
    \cite{Berglund:1995gd}
\begin{equation}
\begin{tabular}{c|c c c c c c c }
~  & $0$ & $1$& $2$& $3 $ & $4$ & $5$ & $6$  \\\hline
$l_1$ & $-2$ & $0$ & $0$& $0$ & $0$ & $1$ & $1$   \\
$l_2$ & $-2$ & $1$& $1$  & $1$& $2$ & $0$ & $-3$
\end{tabular}
\end{equation}

    The mirror manifolds $X^\ast$ is $X^\ast=\widehat{X}/H$, where
$\widehat{X}=X_8(1,1,1,1,4)$ and
$H=(h_i^k)=\frac{1}{8}(7,0,0,1,0),\frac{1}{8}(7,0,1,0,0)$ which act
by $x_i\rightarrow x_i \text{exp}(2\pi ih_i^k)$.

   We consider the following curves
  \bqa
 C_{\alpha,\pm}=\{x_2=\xi
  x_1,x_3=\xi
  x_4,x_5^{2}+\psi x_1x_2x_3x_4x_5=0\}, ~\xi^{8}=-1
  \eqa
which are on the family of divisor  \bqa \eqlabel{Divisor1}
Q(\mathcal{D})=x_2^{8}+z_3x_1^{8}\eqa at the critical points
$z_3=1$.

By the GKZ system, the period on the hypersurface has the form \bqa
\pi=\frac{c}{2}B_{\{\hat{l}_1,\hat{l}_2\}}(u_1,u_2;0,\frac{1}{2})=-\frac{4c}{\pi^{\frac{3}{2}}}u_1\sqrt{u_2}+\mathcal{O}((u_1u_2)^{3/2})
\eqa which vanishes at the critical locus $u_2=0$ expressed in terms
of new parameters as \bqa u_1=z_1 \qquad
u_2=\frac{-z_2}{z_3}(1-z_3)^2 \eqa. Similarly, the off-shell
superpotentials can be obtained by integrating the $\pi$:
  \bqa
  \mathcal{T}_a^{\pm}(z_1,z_2,z_3)=\frac{1}{2\pi
  i}\int\pi(\hat{z})\frac{d\hat{z}}{\hat{z}},
\eqa with the appropriate integral constants\cite{Alim:2010za},
 the superpotentials can be chosen as  $\mathcal{W}^+=-\mathcal{W}^-$. In this convention, the on-shell
 superpotentials can be obtained as
\bqa 2\mathcal{W}^{+}=\frac{1}{2\pi
 i}\int_{-z_3}^{z_3}\pi(z_3)\frac{dz_3}{z_3},~~~W^{\pm}(z_1,z_2)=\mathcal{W}^{\pm}(z_1,z_2,z_3)|_{z_3=1}
\eqa
 Eventually, The superpotential are
\bqa
\begin{split}
&\mathcal{W}^{\pm}(z_1,z_2,z_3)=\mp\sum_{n_1,n_2}\frac{cz_1^{n_1}z_2^{\frac{1}{2}+n_2}z_3^{\frac{-1-2n_1}{2}}\Gamma(2n_1+2n_2+1)}{4\pi(-1+4n_1^2)\Gamma(2+2n_1^2)\Gamma(1+n_2)\Gamma(-\frac{1}{2}-3n_1+n_2)\Gamma(\frac{3}{2}+n_2)}\\
&\{(1-2n_1)
2F_1(-\frac{1}{2}-n_1,-2n_1,\frac{1}{2}-n_1;z_3)+(1+2n_1)z_32F_1((\frac{1}{2}-n_1,-2n_1,\frac{3}{2}-n_1;z_3))\}
\end{split}
\eqa and those can divide two parts \bqa
\mathcal{W}^{\pm}(z_1,z_2,z_3)=W^{\pm}(z_1,z_2)+f(z_1,z_2,z_3)\eqa
where the $f(z_3)$ are related to the open-string parameter, $W$ are
the on-shell superpotential defined as \bqa
W^{\pm}(z_1,z_2=\mp\frac{c}{8}B_{[l_1,l_2]}((z_1,z_2);0,\frac{1}{2})\eqa
 substituting the vector $l_1,l_2$ in this hypersurface, the
 superpotentials are
 \bqa
W^{\pm}=\mp\frac{c}{8}\sum_{n_1,n_2}\frac{z_1^{n_1}z_2^{\frac{1}{2}+n_2}\Gamma(2n_1+2(n_2+\frac{1}{2})+1)}{\Gamma(n_2+\frac{1}{2}+1)^3\Gamma(n_1+1)\Gamma(2(n_1+\frac{1}{2})+1)\Gamma(n_1-3(n_2+\frac{1}{2})+1)}
 \eqa

The additional GLSM charge vectors corresponding to the divisor
\eqref{Divisor1}
   are
       \begin{equation}
\begin{tabular}{c|c c c c c c c c c c}
 & $0$ & $1$& $2$& $3 $ & $4$ & $5$ & $6$ & $7$& $8$&\\\hline
$l_4$ & $0$ & $-1$ & $1$& $0$ & $0$ & $0$ & $0$ & $1$ & $-1$
\end{tabular}
\end{equation}
  The classic A-brane in the mirror Calabi-Yau manifold $X^{\ast}$ of X
determined by the additional charge vectors $(0,-1,1,0,0,0,0)$ is a
special Lagrangian submanifold of $X^{\ast}$ defined as
\cite{Aganagic:2000gs,Aganagic:2001nx,Kachru:2000an,Kachru:2000ih,Alim:2009rf}
\bqa -|x_1|^2+|x_2|^2=\eta\eqa where $x_i$ are coordinates on
$X^{\ast}$, $\eta$ is a K$\ddot{a}$hler moduli parameter with
$\hat{z}=\epsilon e^{-\eta}$ for a phase $\epsilon$.

   \begin{table}[!h]
\def\temptablewidth{1.0\textwidth}
\begin{center}
\begin{tabular*}{\temptablewidth}{@{\extracolsep{\fill}}c|ccccc}
$d_1\backslash \frac{d_2}{2}$ & 1 &3     &5     &7     \\\hline
0               & 2   &-2   &10     &-84     \\
  1              & -40  & 40  &-360& 7232     \\
   2              & -180 &-200  &5500&-215356      \\
   3              &-40 &-29720 &66963200&2314120    \\
   4              & -628&-625424&25006400960 &--
       \end{tabular*}
       {\rule{\temptablewidth}{1pt}}
\end{center}
       \end{table}

    The flat coordinates in
 A-model at large radius regime are related to the flat coordinates of B-model at large complex structure regime by mirror map $t_i=\frac{\omega_i}{\omega_0},~\omega_i:= D_i^{(1)}
 \omega_0(z,\rho)|_{\rho=0}$
 \bqa
 \begin{split}
 &q_1=z_1+2z_1^2+5z_1^3
+2z_1z_2-12z_1^{12}z_2-13z_1z_2^2\mathcal{O}(z^4)\\
 &q_2=z_2-6z_2^2+12z_1z_2+100z_1^2z_2-24z_1z_2^2+\mathcal{O}(z^4)
\end{split}
 \eqa

and we can obtain the inverse mirror map in terms of $q_i=e^{2\pi i
t_i}$
 \bqa
\begin{split}
&z_1=q_1-2q_1^2+3q_1^3-2q_1q_2+48q_1^2q_2+5q_1q_2^2+\mathcal{O}(q^4)\\
 &z_2=q_2+6q_2^2+9q_2^3-12q_1q_2+68q_1^2q_2-168q_1q_2^2\mathcal{O}(q^4).
\end{split}
 \eqa

 Using the modified multi-cover
formula\cite{Walcher:2006rs,Walcher:2009uj} for this case
  \bqa
\frac{W^{\pm}(z(q))}{w_0(z(q))}=\frac{1}{(2 i\pi)^2}\sum_{k~
 odd}~\sum_{d_2 , d_1 odd\geq
0}n_{d_1,d_2}^{\pm}\frac{q_1^{kd_1}q_2^{kd_2/2}}{k^2}\eqa the
superpotentials $W^\pm$, at the critical points $\hat{z}=1$, give
Ooguri-Vafa invariants $n_{d_1,d_2}$ for the normalization constants
  $c=1$, which are listed in Table.1.

  In the limit $Z_1=0$ we also give the superpotential of
  non-compact manifold $\mathcal{O}(-3)_{\mathbb{P}^2}$ which
  can be proved by $n_{0,d_2}=n_k$.
\section{Summary}
~~~For two compact Calabi-Yau threefolds of non-fermat type  constructed by using the toric geometry\cite{Berglund:1995gd},
we study  superpotentials of the D-branes on the these manifolds by extending the GKZ method to the Calabi-Yau manifolds of the non-Fermat type on the B-model side .
 Furthermore, with the mirror symmetry, we calculate the non-perturbative superpotential on the A-model side and extract the Ooguri-Vafa invariants for D-brane
with a single open deformation moduli.

~~~On the other hand, we will study these problems in this paper from an alternative approach \cite{Grimm:2008dq,Grimm:2009sy,Grimm:2010gk,Klevers:2011xs,Grimm:2009ef}
that treats the complex structure deformations of the Calabi-Yau threefolds  and the open string deformations of the D-branes on an equal footing
by studying the complex structure deformations of a non-Calabi-Yau manifold obtained by blowing up the original Calabi-Yau threefold along curve wrapped by the D-brane.
It is interesting to compare the results obtained from the two approaches for the same system of the Calabi-Yau threefold and D-branes.

~~~Furthermore, we are going to treat the obstruction problem to the deformations  with the quiver gauge theories\cite{AF:2005qu} and
the $A_{\infty}$ -algebraic structure \cite{ADDF:2004od, ADD:2002fb, MH:2006fb, CDR:2011mf, AK:2004su, GT:2007su} in the derived category of coherent sheaves.

\section*{Acknowledgments}
~~~~The work is supported by the NSFC (11075204) and President Fund
of GUCAS (Y05101CY00).



\begin{thebibliography}{10}

\bibitem{Candelas}
P.~Candelas, X.~C.~DeLaOssa, P.~S.~Green, and L.Parkes, {A pair of
  Calabi--Yau manifolds as an exactly soluble superconformal theory}, Nucl.
  Phys. B {\bf 359} (1991) 21--74.

\bibitem{Bershadsky:1993cx}
  M.~Bershadsky, S.~Cecotti, H.~Ooguri and C.~Vafa,
  {Kodaira-Spencer theory of gravity and exact results for quantum string amplitudes},
  Commun.\ Math.\ Phys.\  {\bf 165}, 311 (1994)
  [hep-th/9309140].

\bibitem{Aganagic:2000gs}
  M.~Aganagic and C.~Vafa,
 {Mirror symmetry, D-branes and counting holomorphic discs},
  hep-th/0012041.


\bibitem{Aganagic:2001nx}
  M.~Aganagic, A.~Klemm and C.~Vafa,
  {Disk instantons, mirror symmetry and the duality web},
  Z.\ Naturforsch.\ A {\bf 57}, 1 (2002)
  [hep-th/0105045].

\bibitem{Lerche:2002yw}
  W.~Lerche, P.~Mayr and N.~Warner,
 {N=1 special geometry, mixed Hodge variations and toric geometry},
  hep-th/0208039.

\bibitem{Lerche:2002ck}
  W.~Lerche, P.~Mayr and N.~Warner,
 Holomorphic N=1 special geometry of open - closed type II strings,
  hep-th/0207259.

\bibitem{Aganagic:2003db}
  M.~Aganagic, A.~Klemm, M.~Marino and C.~Vafa,
  The Topological vertex,
  Commun.\ Math.\ Phys.\  {\bf 254}, 425 (2005)
  [hep-th/0305132].
\bibitem{Kachru:2000ih}
  S.~Kachru, S.~H.~Katz, A.~E.~Lawrence and J.~McGreevy,
  Open string instantons and superpotentials,
  Phys.\ Rev.\ D {\bf 62}, 026001 (2000)
  [hep-th/9912151].

\bibitem{Kachru:2000an}
  S.~Kachru, S.~H.~Katz, A.~E.~Lawrence and J.~McGreevy,
  Mirror symmetry for open strings,
  Phys.\ Rev.\ D {\bf 62}, 126005 (2000)
  [hep-th/0006047].

\bibitem{Walcher:2006rs}
  J.~Walcher,
  Opening mirror symmetry on the quintic,
  Commun.\ Math.\ Phys.\  {\bf 276}, 671 (2007)
  [hep-th/0605162].
\bibitem{Morrison:2007bm}
  D.~R.~Morrison and J.~Walcher,
  D-branes and Normal Functions,
  arXiv:0709.4028 [hep-th].

\bibitem{Knapp:2008uw}
  J.~Knapp and E.~Scheidegger,
  Towards Open String Mirror Symmetry for One-Parameter Calabi-Yau Hypersurfaces,
  arXiv:0805.1013 [hep-th].

\bibitem{Krefl:2008sj}
  D.~Krefl and J.~Walcher,
  Real Mirror Symmetry for One-parameter Hypersurfaces,
  JHEP {\bf 0809}, 031 (2008)
  [arXiv:0805.0792 [hep-th]].

\bibitem{Jockers:2008pe}
  H.~Jockers and M.~Soroush,
  Effective superpotentials for compact D5-brane Calabi-Yau geometries,
  Commun.\ Math.\ Phys.\  {\bf 290}, 249 (2009)
  [arXiv:0808.0761 [hep-th]].

\bibitem{Grimm:2008dq}
  T.~W.~Grimm, T.~-W.~Ha, A.~Klemm and D.~Klevers,
  The D5-brane effective action and superpotential in N=1 compactifications,
  Nucl.\ Phys.\ B {\bf 816}, 139 (2009)
  [arXiv:0811.2996 [hep-th]].

\bibitem{Grimm:2009sy}
  T.~W.~Grimm, T.~-W.~Ha, A.~Klemm and D.~Klevers,
  Five-Brane Superpotentials and Heterotic / F-theory Duality,
  Nucl.\ Phys.\ B {\bf 838}, 458 (2010)
  [arXiv:0912.3250 [hep-th]].

\bibitem{Alim:2009rf}
  M.~Alim, M.~Hecht, P.~Mayr and A.~Mertens,
  Mirror Symmetry for Toric Branes on Compact Hypersurfaces,
  JHEP {\bf 0909}, 126 (2009)
  [arXiv:0901.2937 [hep-th]].

\bibitem{Alim:2009bx}
  M.~Alim, M.~Hecht, H.~Jockers, P.~Mayr, A.~Mertens and M.~Soroush,
  Hints for Off-Shell Mirror Symmetry in type II/F-theory Compactifications,
  Nucl.\ Phys.\ B {\bf 841}, 303 (2010)
  [arXiv:0909.1842 [hep-th]].

\bibitem{Grimm:2009ef}
  T.~W.~Grimm, T.~-W.~Ha, A.~Klemm and D.~Klevers,
 Computing Brane and Flux Superpotentials in F-theory Compactifications,
  JHEP {\bf 1004}, 015 (2010)
  [arXiv:0909.2025 [hep-th]].

\bibitem{Jockers:2009ti}
  H.~Jockers, P.~Mayr and J.~Walcher,
  On N=1 4d Effective Couplings for F-theory and Heterotic Vacua,
  Adv.\ Theor.\ Math.\ Phys.\  {\bf 14}, 1433 (2010)
  [arXiv:0912.3265 [hep-th]].

\bibitem{Jockers:2009mn}
  H.~Jockers and M.~Soroush,
  Relative periods and open-string integer invariants for a compact Calabi-Yau hypersurface,
  Nucl.\ Phys.\ B {\bf 821}, 535 (2009)
  [arXiv:0904.4674 [hep-th]].

\bibitem{Li:2009dz}
  S.~Li, B.~H.~Lian and S.~-T.~Yau,
  Picard-Fuchs Equations for Relative Periods and Abel-Jacobi Map for Calabi-Yau Hypersurfaces,
  arXiv:0910.4215 [math.AG].

\bibitem{Walcher:2009uj}
  J.~Walcher,
 Calculations for Mirror Symmetry with D-branes,
  JHEP {\bf 0909}, 129 (2009)
  [arXiv:0904.4905 [hep-th]].

\bibitem{Alim:2010za}
  M.~Alim, M.~Hecht, H.~Jockers, P.~Mayr, A.~Mertens and M.~Soroush,
  Type II/F-theory Superpotentials with Several Deformations and N=1 Mirror Symmetry,
  JHEP {\bf 1106}, 103 (2011)
  [arXiv:1010.0977 [hep-th]].

\bibitem{Grimm:2010gk}
  T.~W.~Grimm, A.~Klemm and D.~Klevers,
  ``Five-Brane Superpotentials, Blow-Up Geometries and SU(3) Structure Manifolds,''
  JHEP {\bf 1105}, 113 (2011)
  [arXiv:1011.6375 [hep-th]].

\bibitem{Alim:2011rp}
  M.~Alim, M.~Hecht, H.~Jockers, P.~Mayr, A.~Mertens and M.~Soroush,
  Flat Connections in Open String Mirror Symmetry,
  arXiv:1110.6522 [hep-th].

\bibitem{Klevers:2011xs}
  D.~Klevers,
 Holomorphic Couplings In Non-Perturbative String Compactifications,
  Fortsch.\ Phys.\  {\bf 60}, 3 (2012)
  [arXiv:1106.6259 [hep-th]].

\bibitem{Baumgartl:2007an}
  M.~Baumgartl, I.~Brunner and M.~R.~Gaberdiel,
 D-brane superpotentials and RG flows on the quintic,
  JHEP {\bf 0707}, 061 (2007)
  [arXiv:0704.2666 [hep-th]].

\bibitem{Baumgartl:2008qp}
  M.~Baumgartl and S.~Wood,
 Moduli Webs and Superpotentials for Five-Branes,
  JHEP {\bf 0906}, 052 (2009)
  [arXiv:0812.3397 [hep-th]].

\bibitem{Baumgartl:2010ad}
  M.~Baumgartl, I.~Brunner and M.~Soroush,
  D-brane Superpotentials: Geometric and Worldsheet Approaches,
  Nucl.\ Phys.\ B {\bf 843}, 602 (2011)
  [arXiv:1007.2447 [hep-th]].
\bibitem{Fuji:2010uq}
  H.~Fuji, S.~Nakayama, M.~Shimizu and H.~Suzuki,
 A Note on Computations of D-brane Superpotential,
  J.\ Phys.\ A A {\bf 44}, 465401 (2011)
  [arXiv:1011.2347 [hep-th]].

\bibitem{Shimizu:2010us}
  M.~Shimizu and H.~Suzuki,
 Open mirror symmetry for Pfaffian Calabi-Yau 3-folds,
  JHEP {\bf 1103}, 083 (2011)
  [arXiv:1011.2350 [hep-th]].

\bibitem{Feng}
Feng-Jun Xu, Fu-Zhong Yang {Type II/F-theory Superpotentials and
Ooguri-Vafa Invariants of Compact Calabi-Yau Threefolds with Three
Deformations},  [hep-th/1206.0445].
\bibitem{GKZ}
  I.M.Gel'fand, A.Zelevinski and M.Kapranov,
 Funct.Anal.Appl. {\bf 28}, 94 (1989)
  [arXiv:9308122 [hep-th]].

\bibitem{Gukov:1999ya}
  S.~Gukov, C.~Vafa and E.~Witten,
  CFT's from Calabi-Yau four folds,
  Nucl.\ Phys.\ B {\bf 584}, 69 (2000)
  [Erratum-ibid.\ B {\bf 608}, 477 (2001)]
  [hep-th/9906070].

\bibitem{Berglund:2005dm}
  P.~Berglund and P.~Mayr,
  Non-perturbative superpotentials in F-theory and string duality,
  hep-th/0504058.

\bibitem{Hosono:1993qy}
  S.~Hosono, A.~Klemm, S.~Theisen and S.~-T.~Yau,
  Mirror symmetry, mirror map and applications to Calabi-Yau hypersurfaces,
  Commun.\ Math.\ Phys.\  {\bf 167}, 301 (1995)
  [hep-th/9308122].

\bibitem{Hosono:1995bm}
  S.~Hosono, B.~H.~Lian and S.~T.~Yau,
  GKZ generalized hypergeometric systems in mirror symmetry of Calabi-Yau hypersurfaces,
  Commun.\ Math.\ Phys.\  {\bf 182}, 535 (1996)
  [alg-geom/9511001].

\bibitem{Hosono}
  S.~Hosono, B.~H.~Lian and S.~T.~Yau,
 GKZ Systems, Grobner Fans and Moduli Spaces of Calabi-Yau Hypersurfaces,
[arXiv:alg-geom/9707003v2].

\bibitem{Batyrev:1993wa}
  V.~V.~Batyrev and D.~van Straten,
 Generalized hypergeometric functions and rational curves on Calabi-Yau complete intersections in toric varieties,
  Commun.\ Math.\ Phys.\  {\bf 168}, 493 (1995)
  [alg-geom/9307010].

\bibitem{Batyrev:1994hm}
  V.~V.~Batyrev,
  Dual polyhedra and mirror symmetry for Calabi-Yau hypersurfaces in toric varieties,
  J.\ Alg.\ Geom.\  {\bf 3}, 493 (1994)
  [alg-geom/9310003].


\bibitem{Mayr:2001xk}
  P.~Mayr,
  N=1 mirror symmetry and open / closed string duality,
  Adv.\ Theor.\ Math.\ Phys.\  {\bf 5}, 213 (2002)
  [hep-th/0108229].


\bibitem{Witten:1992fb}
  E.~Witten,
 Chern-Simons gauge theory as a string theory,
  Prog.\ Math.\  {\bf 133}, 637 (1995)
  [hep-th/9207094].


\bibitem{Lerche:2003hs}
  W.~Lerche,
  Special geometry and mirror symmetry for open string backgrounds with N = 1 supersymmetry,
  hep-th/0312326.




\bibitem{Mayr:2000hh}
  P.~Mayr,
 {On supersymmetry breaking in string theory and its realization in brane worlds},
  Nucl.\ Phys.\ B {\bf 593}, 99 (2001)
  [hep-th/0003198].

\bibitem{Taylor:1999ii}
  T.~R.~Taylor and C.~Vafa,
  RR flux on Calabi-Yau and partial supersymmetry breaking,
  Phys.\ Lett.\ B {\bf 474}, 130 (2000)
  [hep-th/9912152].


\bibitem{Witten:1997ep}
  E.~Witten,
 Branes and the dynamics of QCD,
  Nucl.\ Phys.\ B {\bf 507}, 658 (1997)
  [hep-th/9706109].


    \bibitem{Clemens}
 H.~Clemens,
 Cohomology and Obstructions II: Curves on K-trivial threefolds,
  arXiv:math/0206219.



\bibitem{Griffiths}
P.~Griffiths, A theorem concering the differential equations
satisfied by normal functions associated to algebraic cycles,
  Am.\ J.\ Math.\  {\bf 101}, 96 (1979)

\bibitem{Ooguri:1999bv}
  H.~Ooguri and C.~Vafa,
  Knot invariants and topological strings,
  Nucl.\ Phys.\ B {\bf 577}, 419 (2000)
  [hep-th/9912123].

\bibitem{Fujino}
 O.~Fujino and H.~Sato,
 Introduction to the toric Mori theory,
 arXiv:math/0307180v2 [math.AG]

\bibitem{Fujino2}
 O.~Fujino,
 Notes on toric varieties from Mori theoretic viewpoint,
    arXiv:math/0112090v1 [math.AG]


\bibitem{Scaramuzza}
A.~Scaramuzza, Smooth complete toric varieties: an algorithmic
approach. Ph.D. dissertation, University of Roma Tre, 2007.

\bibitem{Renesse}
C.~V.~Renesse, Combinatiorial aspects of toric varieties. Ph.D.
dissertation, University of Massachusetts Amherst, 2007.

\bibitem{Witten:1993yc}
  E.~Witten,
  Phases of N=2 theories in two-dimensions,
  Nucl.\ Phys.\ B {\bf 403}, 159 (1993)
  [hep-th/9301042].

\bibitem{Berglund:1998ej}
  P.~Berglund and P.~Mayr,
  Heterotic string/F theory duality from mirror symmetry,
  Adv.\ Theor.\ Math.\ Phys.\  {\bf 2}, 1307 (1999)
  [hep-th/9811217].



\bibitem{Berglund:1995gd}
  P.~Berglund, S.~H.~Katz and A.~Klemm,
  Mirror symmetry and the moduli space for generic hypersurfaces in toric varieties,
  Nucl.\ Phys.\ B {\bf 456}, 153 (1995)
  [hep-th/9506091].

\bibitem{Lerche:1996ni}
  W.~Lerche, P.~Mayr and N.~P.~Warner,
  Noncritical strings, Del Pezzo singularities and Seiberg-Witten curves,
  Nucl.\ Phys.\ B {\bf 499}, 125 (1997)
  [hep-th/9612085].


\bibitem{AF:2005qu}
Paul S. Aspinwall, Lukasz M. Fidkowski,
Superpotentials for Quiver Gauge Theories,
JHEP 0610:047,(2006)
[hep-th/0506041].


\bibitem{ADDF:2004od}
Sujay K. Ashok, Eleonora Dell'Aquila, Duiliu-Emanuel Diaconescu, Bogdan Florea,
Obstructed D-Branes in Landau-Ginzburg Orbifolds,
Adv. Theor. Math.  Phys. {\bf 8}  427-472 (2004)
[hep-th/0404167].
\bibitem{ADD:2002fb}
Sujay K. Ashok, Eleonora Dell'Aquila, Duiliu-Emanuel Diaconescu,
Fractional Branes in Landau-Ginzburg Orbifolds ,
Adv.\ Theor.\ Math.\ Phys. {\bf 8}  461-513 (2004)
[hep-th/0401135].
\bibitem{MH:2006fb}
Manfred Herbst,
Quantum A-infinity Structures for Open-Closed Topological Strings ,
[hep-th/0602018].
\bibitem{CDR:2011mf}
Nils Carqueville, Laura Dowdy, Andreas Recknagel ,
Algorithmic deformation of matrix factorisations ,
JHEP 04 (2012) 014
[1112.3352].
\bibitem{AK:2004su}
Paul S. Aspinwall, Sheldon Katz,
Computation of Superpotentials for D-Branes
Commun.Math.Phys. 264, 227-253  (2006)
[hep-th/0412209].
\bibitem{GT:2007su}
Gueorgui Todorov,
D-branes, obstructed curves, and minimal model superpotentials,
[0709.4673].

\end{thebibliography}
\addcontentsline{toc}{section}{References}

\begingroup\raggedright\endgroup

\end{document}